# Bonding trends within ternary Isocoordinate chalcogenide glasses $Ge_xAs_ySe_{1-x-y}$

G. Opletal
Department of Applied Physics, School of Applied Sciences, RMIT University.
124 La Trobe Street, Melbourne, Victoria 3000, Australia.

R. P. Wang
Centre for Ultrahigh Bandwidth Devices for Optical Systems (CUDOS), Laser Physics Centre,
Research School of Physics and Engineering, The Australian National University.
Canberra, ACT 0200, Australia

S. P. Russo
Department of Applied Physics, School of Applied Sciences, RMIT University.
124 La Trobe Street, Melbourne, Victoria 3000, Australia.

A structural study is presented of *ab-initio* molecular dynamics simulations of Ge-As-Se calcogenide glasses performed at the same mean coordination number but differing stoichiometry ranging between Se rich and Se poor glasses. Starting configurations are generated via Reverse Monte Carlo (RMC) simulations of Extended X-ray Absorption Fine Structure (EXAFS) measurements of experimental samples. Structural analysis is presented illustrating the bonding trends found with changing stoichiometry.

I. Introduction

With ultrafast broad-band response times and optical nonlinearity, chalcogenide glasses have received considerable research interest over the past few decades holding promising potential as photonic devices.[1,2,3] Ge-As-Se systems in particular are important due to their wide glass formation region and much attention has been spent attempting to develop an understanding of their non crystalline microstructure.

Thorpe *et al*.[4] has shown that the GeAsSe system have three phases with differing microstructure dependent upon the mean coordination number (MCN). The MCN is calculated under the assumption that the Ge, As and Se fractions only contain four, three and two bonds respectively. Experiments have illustrated that many properties have been found in Ge-As-Se systems to depend only upon the MCN and not the detailed stoichiometry of the glass.[5] At low MCN, the so called 'floppy' phase is thought to consist of Ge-Se4 tetrahedral and As-Se3 pyramid units with longer chains of Se-Se atoms resulting from an excess of Se atoms. At a MCN of 2.4, a transition occurs to a 'rigid' phase where there is less Se atoms available for Se chains beyond the saturation of the Ge and As bonds. As the MCN increases further, a lack of Se atoms, leads to increasing amounts of As-As, Ge-Ge and As-Ge bonding as observed via Raman spectroscopy.[6] An intermediate thermally reversible phase can also exist between the floppy and rigid phases whose width has been linked to structural variability and chemical composition.[7]

Whilst experimental techniques provide clues about the microstructure, computer simulations offer complete structure information in the form of atomistic structural models derived either from molecular dynamics (MD) or some other Metropolis Monte Carlo (MC) based schemes. Chalcogenide glass systems have been modeled in the past by MC based methods which incorporate a fitting to experiment diffraction data[8,9] and more recently, ab-*initio* MD simulations.[10,11,12] In particular, for the Ge-As-Se systems, the ab-*intio* MD simulations of Cai *et al.*[11] are particularly relevant. Here the authors describe a fundamental concern in regards to the accuracy of such models when using traditional quench based MD simulations over short time scales starting from a random starting structure, as is typically the case in computationally intensive first principle methods. Irrespective of the accuracy of the Hamiltonian employed, if there is insufficient configurational sampling, the resulting models risk getting trapped in local energy minima. The authors approach this issue by employing a method of producing their larger system via multiple replicas of smaller relaxed 'building block' cells.

In this work, we develop five models for Ge-As-Se systems of different stoichiometry but with the same MCN and approach the random starting structure issue by using initial configurations derived from Reverse Monte Carlo (RMC)[13] simulations. The five systems are chosen to range from Se poor to Se rich environments and thus investigate the degree of structural variability and bonding trends that can occur in the glasses by altering their chemical composition.

**II. Methodology**

Five glass samples were manufactured with stoichiometric compositions that give a theoretical MCN of 2.5 atoms which is calculated by $4N_{GE} + 3N_{AS} + 2(1-N_{GE}-N_{AS})$ where $N_{GE}$ and $N_{AS}$ is the fraction of Ge and As atoms respectively.  Under such conditions where each Se atom only bonds to four Ge atoms and three As atoms, the expected Se concentration to saturate all the Ge and As bonds is $2N_{GE} + 1.5N_{AS}$ due to the sharing of the Se atoms.  The samples are listed in table I in increasing Ge and Se concentration along with the density, number of atoms in simulation cell and the Se content.

The method of generation of the initial structures was motivated by the past reported difficulties of obtaining realistic models via quenched MD simulations when starting from a random starting structure.[11,14]  The RMC methodology was used which fit experimentally measured Ge, As and Se K-edge Extended X-ray Absorption Fine Structure (EXAFS) measurements and other experimentally motivated constraints.

The resulting RMC models were then employed as initial starting structures for ab-*initio* molecular dynamics simulations using the package Vienna Ab-*initio* Simulation Package (VASP).[15]  Following a temperature quench to 300K, structural analysis was performed over an ensemble of configurations.

A.      **Experimental Details**

Chalcogenide glasses were prepared from 50 g batches of high purity (5N) Germanium, Arsenic and Selenium metals. The required amounts of these raw materials were weighed inside a dry nitrogen glove box and loaded into a pre-cleaned quartz ampoule. The loaded ampoule was dried under vacuum (10–6 Torr) at 110°C for 4 hours to ensure removal of surface moisture from the raw materials. The ampoule was then sealed under vacuum using an oxygen hydrogen torch, and introduced into a rocking furnace for melting of the contents at 900°C. The melt was homogenized for a period not less than 30 hours, then the ampoule was removed from the rocking furnace at a predetermined temperature and air quenched. The resulting glass boule was subsequently annealed at a temperature 30°C below its glass transition temperature Tg, then slowly cooled to room temperature.  Following the annealing process, the glass boules were sectioned to form discs of 25 mm diameter and approximately 2 mm thick.  The density of the samples was measured using a Mettler $H_2O$ balance (Mettler-Toledo Ltd., Switzerland) with a-MgO crystal used as a reference. Samples from each glass composition were weighed five times in air and ethanol, respectively, and the average density was recorded. The error bar for each measurement being less than 0.005 $gcm^{-3}$.

Ge, As and Se K-edge EXAFS measurements were conducted on all five sample in transmission mode.  From the resulting X-ray absorption cross-sections µ(E), the EXAFS signals χ(k) were extracted using the data processing program Visual Processing for EXAFS Researches (VIPER)[16].  These signals were Fourier transformed and the first nearest neighbor contribution peak was extracted by filtering using a Hanning window function before transforming back to r-space. This procedure has been employed previously for data preparation in previous RMC studies[17,18].

B.     Simulation Details

In order to produces initial configurations consistent with the EXAFS datasets, the RMC methodology [13] was employed. Briefly, the RMC method is a variation of the Metropolis-Hasting algorithm [19,20] and attempts to produce coordinate configurations consistent with experimental data such as a pair distribution functions, structure factors or EXAFS data. In a typical single move RMC simulation, a random atom is chosen and randomly displaced. If the fit to the experimental data is improved, the move is accepted. Otherwise, the move is rejected probabilistically. The simulation is left to evolve until the discrepancy between the simulation and experimental data oscillates about some value. Explicit details are given in McGreevy.[21]

For this work, the RMC++ code[22] was employed since it can fit EXAFS datasets for multi-component systems alongside other constraint such as bond angles. In order to calculate the model EXAFS $\chi(k)$ functions within RMC++, the EXAFS backscattering amplitudes were calculated using the ab-*initio* multiple scattering FEFF 9.03 code[23].

Thus, in addition to the experimental stoichiometry and density, three experimentally motivated constraints were used within the RMC simulations. EXAFS $\chi(k)$ signal constraints for Ge, As and Se with a relative medium weighting. A coordination constraint attempting to achieve four neighbors per atom for Ge, three for As and two for Se with a low relative weighting and an exclusion of angles around 60 degrees in the bond angle distribution $g_3(\theta)$ with a high relative weighting.  The last constraint was required to avoid the creation of large three member ring populations which can occur in under-constrained RMC simulations.[24]

Of key importance within the simulations was the allowance for switching moves alongside the traditional translation moves.  Switching moves simply switch the labels of two selected different elements and aid in configuration sampling in multi-component systems.   It should be noted that the RMC models themselves are not expected to be accurate representations of the real systems as it is almost certain that they are 'under-constrained' for such a complicated system.  They are however considered useful in reducing the amount of configuration space that needs to be sampled in the subsequent computationally expensive MD simulations.  In particular, the EXAFS datasets provide information on elemental first nearest neighbor coordination.  The resulting RMC models where then used as the initial starting configuration for subsequent molecular dynamics simulations using VASP.

VASP[15] is an ab-*initio* quantum mechanical code utilizing a plane wave basis set within a finite temperature local density approximation (LDA) that can perform molecular dynamics by an evaluation at each ion steps of the electronic ground state.  Sampling was restricted to the Γ point in the Brillouin zone and an ionic time step of 3 fs was used along with a Nosé thermostat.  The structures were equilibrated at 2000K over 20ps and then quenched linearly to 300K over a further 30ps.  After a further 5ps equilibration at 300K, the final configurations were collected and analyzed using a variety of structure measures.

Structural measures such as pair correlation functions g(r), bond angle distributions $g_3(r)$, network ring statistics and bonding environments were averaged over 500 configurations or 1.5 ps at 300K to obtain adequate statistics.  This was performed because of fluctuations in MCN and other quantities at room temperature.  A cut-off distance of 2.85 Å was used to define the first nearest neighbors which was found to be the minimum in the total and partial g(r) curves between the first and second peaks.   Ring statistics, calculated for shortest path rings using the method of Franzblau[25], were terminated at six members to avoid counting contribution found through the periodic boundaries.

## III. Results and Discussion

A variety of structure measure averages are illustrated in table II. Firstly, it can be seen that all of the configurations resulted in an average MCN of approximately 2.5 irrespective of the changes to the stoichiometry. The As and Se average coordination remained roughly static over the models while the Ge average coordination increased with increasing Se concentration. Based upon bond energy arguments[26], it has been suggested that Se atoms would first saturate the Ge bonds prior to As atoms. However, in this work, we see the As atoms having three nearest neighbors almost exclusively across all five systems while it is the Ge atoms that vary and generally increase in coordination with increase Se and Ge concentration.

Also visible is a significant fraction of over-coordinated Se atoms in addition to Ge atoms having both three and five nearest neighbors. Various non-ideal bonds are also seen in particular in the Se poor systems with the appearance of As-As, Ge-Ge and Ge-As bonds, all of which is consistent with the models of Cai *et al.*[11] A small number of Se atoms having a single nearest neighbor also appear in the averages and is partly due to the fact that the averaging was performed near room temperature without any final geometry optimization and in the presence of the inherent thermal fluctuations induced by Nosé thermostat.

A peculiar observation is seen in the 15_20_65 system which has a reduction in fourfold Ge coordination contrary to the increasing trend with increasing Ge concentration. This is due to a spike in the Ge-As3 population while the Ge-Se4 fraction remains unchanged when comparing systems 15_20_65 and 12.5_25_62.5. A closer look in Fig. **1** at the fraction of fourfold Ge atoms being in the Ge-Se4 configuration illustrates that the 15_20_65 system still follows the general trend of moving towards tetrahedral bonding with increasing Se concentration. Figure 1 also illustrates a similar trend for As-Se3 pyramidal units as a fraction of threefold As atoms along with the expected reduction in Ge and As centered bonding units bonded to other Ge and As atoms. An exception is seen in the 7.5_35_57.5 system which has its few Ge atoms bonded to a significant fraction of Se atoms. There is likely a significant variation of these values depending on the environment the few Ge atoms settle in at low concentrations.

An illustration of the bonding in two of the system is shown in Fig. **2**. The Se very poor configuration (7.5_35_57.5) illustrates a structure containing significant numbers of Ge-As and As-As bonds in contrast to the Se very rich configuration (20_10_70).

An illustration of the radial distribution function g(r) is shown in Fig. **3** along with a decomposition of the g(r) into partial g(r) contributions from Ge, As and Se atoms in Fig. **4**. At first sight, the total g(r) curves are similar and in fact closely matches the total g(r) of the $Ge_2As_2Se_4$ system of Cai et al.[11] However, the total g(r) first peak shifts slightly to lower values of r with increasing Se concentration while the second peak shifts to the higher values of r for the 20_10_70 system. A comparison of the partial g(r) curves of the 7.5_35_57.5 and 20_10_70 systems show that the shortening is caused by a decrease in bond length of the Ge-Se bonds and an increase in the number of short Se-Se bond in the Se rich system. The change to larger r in the second peak position is due to an increase in the Se-Se second neighbor distance.

The bond angle distributions for the five systems and some relevant partial distributions are shown in Fig. **5**. A number of observations can be made.

Firstly, a significant shift is observed in the distributions to larger bond angles with increasing Se concentration, in particular between the 15_10_70 and 20_10_70 system. This increase is seen in the average bond angles (97.8, 98.4, 99.7, 99.6, 102.1 degrees for the five systems) with increasing Se concentration and reflects a rise in fourfold tetrahedral units. The primary contribution to this shift is the Se-Ge-Se peak (the labeling representing the angle between a Ge atom and its two Se neighbours) which grows and shifts as the average coordination of Ge increases from 3.40 to 3.58 between the 15_20_65 and 20_10_70 systems (20_10_70 Se-Ge-Se peak in Fig. **4**). This coincides with a decrease in the Se-As-Se peak and represents a shift from three fold pyramidal As-Se3 unitsto Ge-Se4 tetrahedral units. Similar trends were observed in Raman experiments where high Ge concentration and reduced Se concentration reduced the As-Se vibration band which was interpreted as a reduction in As-Se3 units.[6]

A second observation is the appearance of a peak around 60 degrees, representing small three member rings, which tend to decrease in number with increasing Ge and Se concentration. This peak is almost entirely composed of As-As-Se and As-Se-As contributions in the Se poor systems and also includes contribution from Ge-Ge-Se and Ge-Se-Ge in the Se rich systems. This peak was not an artifact of the RMC procedure which itself contained very few such three member rings.

A three member As-Se-As ring can be seen in the 7.5_35_57.5 model illustration in Fig. **2** and can also be observed in the models of Cai *et al.*[11] in their Fig. **2(a)** and thus may represent a realistic environment at significant As concentrations.

Figure **6** illustrates the ring size distribution of the various structures. The decreasing three member ring population can be directly seen to mimic the decrease in the 60 degree peak in the bond angle distribution. In the Se rich systems, we see an increase in larger six member rings although the number of five member rings decreases. Three member rings were also observed in $GeSe_2$ ab-*initio* simulations of Cobb *et al.*[10] along with a similar ring size distribution (reported at 3, 20, 10, 23 rings for 3, 4, 5 and 6 member rings) to the increasingly similar 20_10_70 system.

The dominant bonding environments within all five systems are presented in Fig. **7**. The As-Se3 reduction and growth of Ge-Se4 with increasing Ge and Se concentration can be directly seen. In addition, the most numerous structural unit in the Se poor system, Se-As2 is replaced by Ge-Se-Se in the Se rich system and Se-Se-Se chains begin to appear with increasing Se concentration.

Se chains were not observed in the simulation of Cai *et al.*[11], nor were any Se-Se bonds. This result can be reconciled by looking at the Se-Se homonuclear bond formation fraction in Fig. **8** as a function of Se content.

Assuming the linear relationship holds down to lower Se concentrations, it suggests that homonuclear Se-Se linear bonds disappear at 50% Se concentration which explains why the low Se models (with 31.7% and 40% Se concentration) of Cai et al.[11] never observed any such bonds.

A closer look at the Se-Se bonding is shown in **Fig. 9** where the number of Se chain segments is plotted for the five systems. The Se very poor 7.5_35_57.5 system only contained single Se-Se bonds. Triple Se-Se-Se segments begin to appear in the Se poor 10_30_60 system which is in line with x-ray photoelectron spectroscopy experiments of Wang et al.[27], where trimers were detected even in Se poor samples. With increase Se content, the Se chain length distribution broadens.

## IV. Conclusions

Ab-*initio* MD models of GeAsSe glasses were produced starting from EXAFS constraint RMC structures at MCN =2.5 with varying stochiometry ranging from Se poor to Se rich glasses. Although the simplistic assumption of Ge, As and Se atoms having four, three and two bonds respectively was seen to be incorrect, the systems partial coordinations compensated each other to keep the MCN constant. The Se poor systems contained significant homopolar and non Se bonding, short Se chains and were dominated by Se-As2 and As-Se3 units. With increasing Ge and Se concentration, a reduction of non-Se bonding units was observed and instead was replaced by Ge-Se4 units and Se chains of increasing length. Further studies with varying MCN are underway and should aid in mapping the bonding environments present in these material at different elemental concentrations.

**Acknowledgements**

We are grateful for the funding support from the Australian Research Council (ARC) Centre of Excellence to the Centre for Ultrahigh Bandwidth Devices for Optical Systems (CUDOS) and the ARC discovery project.  In addition, we are thankful to the National Computational Infrastructure (NCI) for the computational resources required for this study.

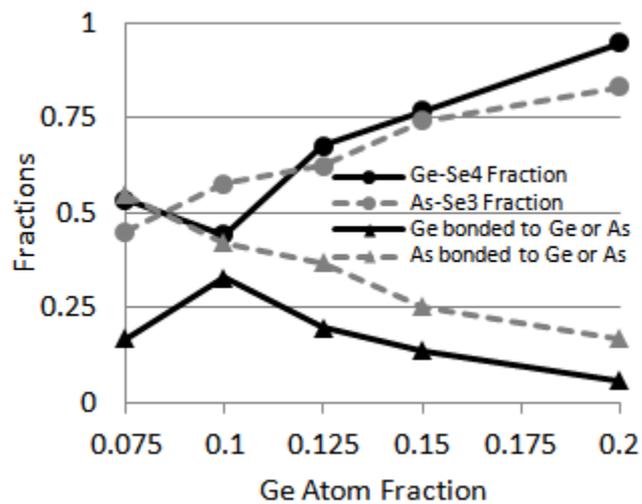

FIG. 1. Fraction of fourfold Ge atoms in the Ge-Se4 configuration, three fold As atoms in the As-Se3 configuration and fraction of Ge and As atoms having at least one bond to other Ge or As atoms.

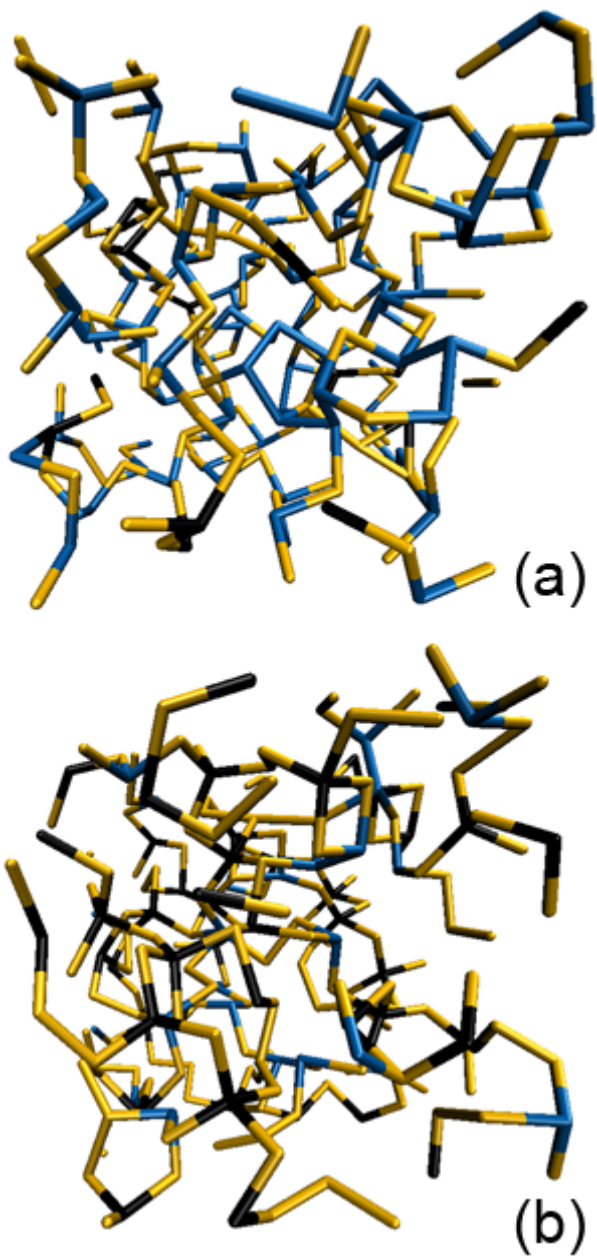

FIG. 2. (Color online) Snapshots of the (a) 7.5_35_57.5 system and (b) 20_10_70 system at 300K. Colour scheme has Ge as black (black), As blue (grey) and Se yellow (light).

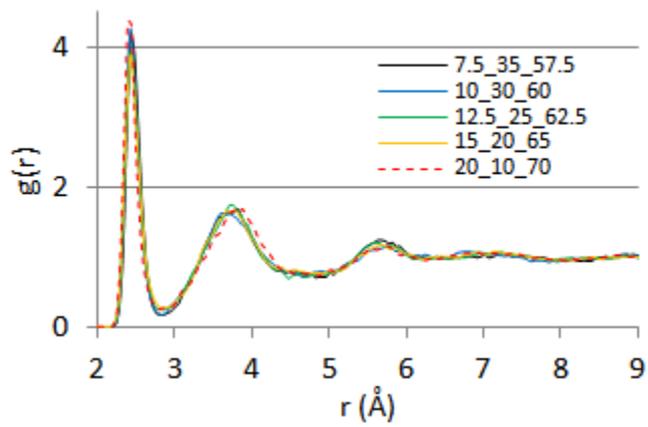

FIG. 3. (Color online) Total g(r) for all five systems. A small shift to shorter distances is observed in the first peak and longer distance in the second peak as a function of increasing Se and Ge concentration.

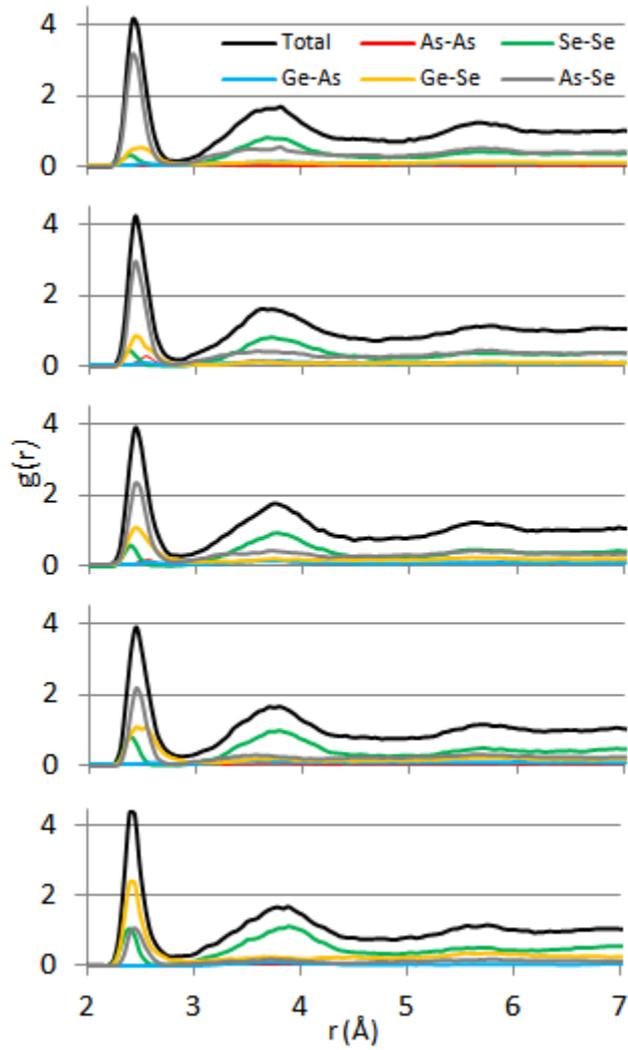

FIG. 4. (Color online) Partial g(r) for the five systems in increasing Ge and Se concentration. Top to bottom are the 7.5_35_57.5, 10_30_60, 12.5_25_62.5, 15_20_65 and 20_10_70 systems.

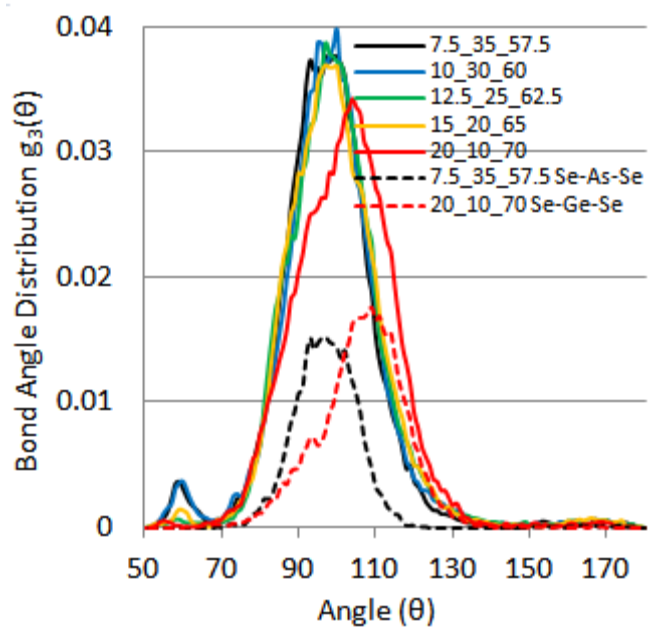

FIG. 5. (Color online) Bond angle distributions for the fives systems along with contributions the dominant partial contributions for system 7.5_35_57.5 and 20_10_70.

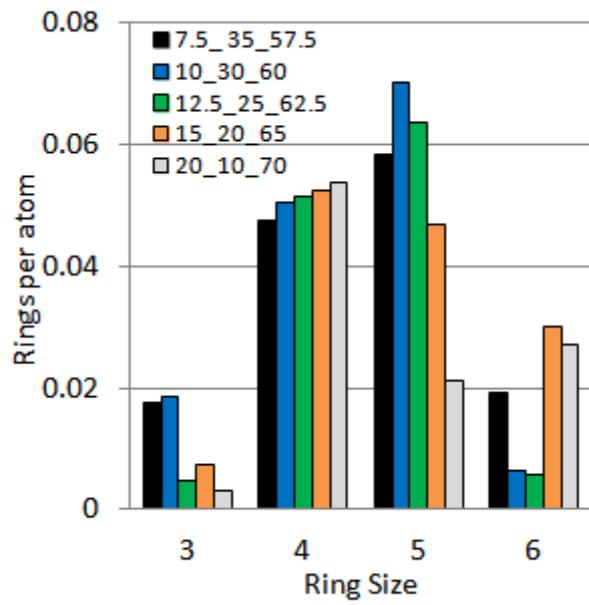

FIG. 6. (Color online) Ring size distributions averaged over 500 configurations for the five systems.

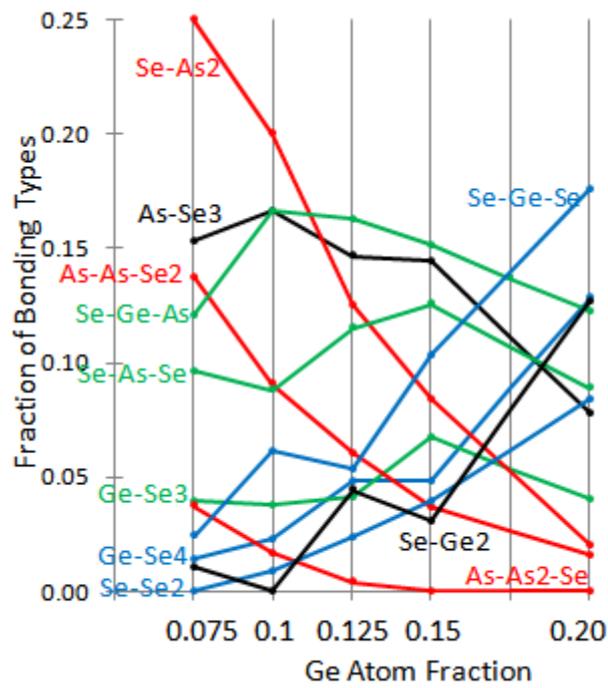

FIG. 7. (Color online) Dominant bonding environments for the five systems. The gap at 0.175 reflects a lack of data at that concentration.

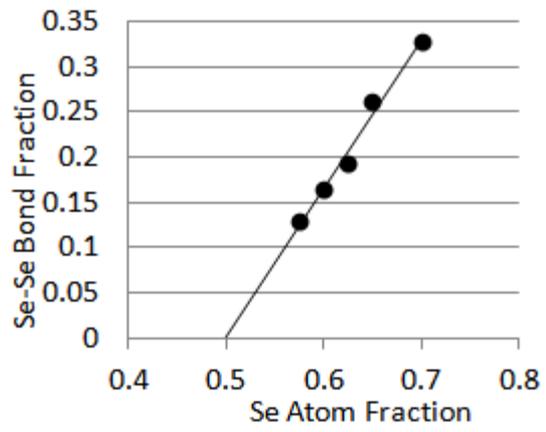

FIG. 8. Fraction of Se-Se bonds versus Se concentration. A linear fit and extrapolation is also shown.

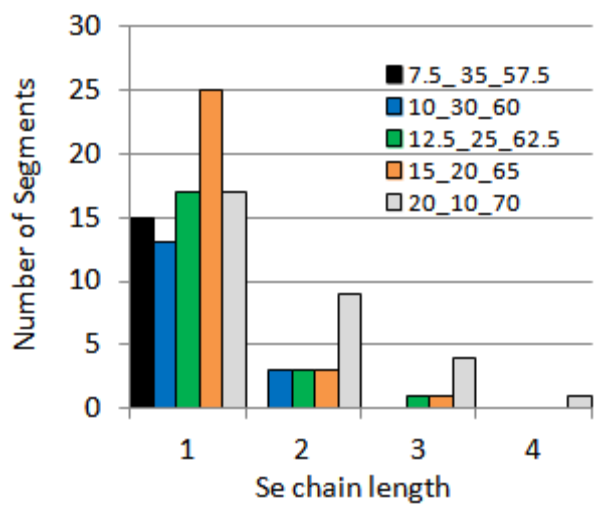

FIG. 9. (Color online) Se-Se chains length populations for the five systems.

| Name | Density (g/cm³) | Ge | As | Se | Se content |
|---|---|---|---|---|---|
| 7.5_35_57.5 | 4.453 | 18 | 84 | 138 | Very Poor (67.5) |
| 10_30_60 | 4.469 | 24 | 72 | 144 | Poor (65) |
| 12.5_25_62.5 | 4.493 | 30 | 60 | 150 | Even (62.5) |
| 15_20_65 | 4.431 | 36 | 48 | 156 | Rich (60) |
| 20_10_70 | 4.412 | 48 | 24 | 168 | Very Rich (55) |

Table I. Sample labels, density and atom numbers and Se contents used within the simulations. The stoichiometric composition fractions are contained within the sample labels and the Se content bracket value is the ideal content assuming all Ge and As bonds are saturated given by $2N_{GE} + 1.5N_{AS}$.

|  | | Coordination | | | | | Bond Type | | |
|---|---|---|---|---|---|---|---|---|---|
|  | Mean | 1 | 2 | 3 | 4 | 5 | Ge | As | Se |
|  | **7.5_35_57.5** | | | | | | | | |
| Ge | 3.26 | 0.0 | 10.7 | 53.1 | 36.1 | 0.1 | 0.0 | 5.2 | 94.8 |
| As | 3.03 | 0.0 | 0.1 | 97.2 | 2.7 | 0.0 | 1.2 | 20.5 | 78.3 |
| Se | 2.12 | 0.4 | 87.3 | 12.2 | 0.0 | 0.0 | 19.0 | 68.1 | 12.9 |
| Mean | 2.52 | 0.2 | 51.1 | 45.0 | 3.7 | 0.0 | | | |
|  | **10_30_60** | | | | | | | | |
| Ge | 3.47 | 0.0 | 5.9 | 41.7 | 51.9 | 0.5 | 6.8 | 4.8 | 88.3 |
| As | 3.03 | 0.0 | 0.6 | 96.3 | 3.0 | 0.1 | 1.9 | 14.9 | 83.3 |
| Se | 2.12 | 0.5 | 87.5 | 11.9 | 0.1 | 0.0 | 24.1 | 59.6 | 16.3 |
| Mean | 2.52 | 0.3 | 53.3 | 40.2 | 6.1 | 0.1 | | | |
|  | **12.5_25_62.5** | | | | | | | | |
| Ge | 3.52 | 0.1 | 8.1 | 33.2 | 57.1 | 1.5 | 1.9 | 5.6 | 92.5 |
| As | 3.03 | 0.0 | 1.5 | 94.2 | 4.3 | 0.0 | 3.3 | 10.0 | 86.7 |
| Se | 2.11 | 2.5 | 84.1 | 13.3 | 0.0 | 0.0 | 30.9 | 49.8 | 19.3 |
| Mean | 2.51 | 1.6 | 53.9 | 36.0 | 8.2 | 0.2 | | | |
|  | **15_20_65** | | | | | | | | |
| Ge | 3.40 | 0.1 | 7.5 | 47.7 | 41.9 | 2.8 | 1.6 | 2.5 | 95.9 |
| As | 3.03 | 0.0 | 0.2 | 97.0 | 2.7 | 0.1 | 2.1 | 6.9 | 91.0 |
| Se | 2.16 | 0.7 | 82.5 | 16.7 | 0.1 | 0.0 | 34.8 | 39.2 | 26.1 |
| Mean | 2.52 | 0.4 | 54.8 | 37.4 | 6.9 | 0.4 | | | |
|  | **20_10_70** | | | | | | | | |
| Ge | 3.58 | 0.3 | 9.6 | 22.0 | 67.8 | 0.3 | 2.0 | 0.0 | 97.9 |
| As | 3.04 | 0.0 | 0.9 | 93.9 | 5.2 | 0.0 | 0.1 | 5.5 | 94.4 |
| Se | 2.10 | 0.8 | 88.5 | 10.5 | 0.2 | 0.0 | 47.7 | 19.5 | 32.7 |
| Mean | 2.49 | 0.6 | 64.0 | 21.1 | 14.2 | 0.1 | | | |

Table II. Averaged analysis of coordination and bonding types. The 'mean' row for each system represents an average over all element types while the 'mean' column represents averages for individual elements. Averages are over 500 configurations (1.5ps) at 300K. Values are percentages.